\documentclass[10pt,journal]{IEEEtran}
\usepackage{lineno,hyperref}
\usepackage{times}
\usepackage{url}
\usepackage{latexsym}
\usepackage{mathrsfs}
\usepackage{xcolor}
\usepackage{microtype}
\usepackage{graphicx}
\usepackage{booktabs}
\usepackage{microtype}
\usepackage{pifont}
\usepackage{latexsym}
\usepackage{float}
\usepackage{multirow}
\usepackage{enumitem}
\usepackage{algorithm}
\usepackage{algpseudocode}
\usepackage{booktabs}
\usepackage{amssymb,mathtools}
\usepackage{eqnarray,amsmath}
\usepackage{hyperref}
\usepackage{microtype}
\usepackage{subfig}
\usepackage{lscape}
\usepackage{arydshln}
\usepackage{orcidlink}
\usepackage{tikz}
\usepackage{cuted}
\usetikzlibrary{positioning}
\modulolinenumbers[5]
\newcolumntype{C}[1]{>{\centering\arraybackslash}m{#1}}

\ifCLASSINFOpdf 
\else 
\fi 

\newcommand{\beginsupplement}{
  \renewcommand{\thesection}{S\arabic{section}}
  \renewcommand{\thesubsection}{S\arabic{section}.\arabic{subsection}}
  \renewcommand{\thesubsubsection}{S\arabic{section}.\arabic{subsection}.\arabic{subsubsection}}
  \setcounter{section}{0}
  \renewcommand{\thefigure}{S\arabic{figure}}
  \renewcommand{\thetable}{S\arabic{table}}
  \renewcommand{\theequation}{S\arabic{equation}}
  \setcounter{figure}{0}\setcounter{table}{0}\setcounter{equation}{0}
}

\makeatletter

\makeatletter

\begin{document}

%% =*=*=*=*=*=*=*=*=*=*=*=*=*=*=*=*=*=*=*=*=*=*=*=*=*=*=*=*=*=*
\title{Synthesizing Sentiment-Controlled Feedback For Multimodal Text and Image Data}

\author{
    Puneet~Kumar
        \orcidlink{0000-0002-4318-1353},
        \IEEEmembership{Member,~IEEE},
    Sarthak~Malik
        \orcidlink{0000-0001-5224-1445},
    Balasubramanian~Raman
        \orcidlink{0000-0001-6277-6267},
        \IEEEmembership{Senior~Member,~IEEE} and 
    Xiaobai Li*
        \orcidlink{0000-0003-4519-7823},
        \IEEEmembership{Senior~Member,~IEEE}\thanks{*Corresponding Author.}
    \thanks{P. Kumar is with the Center for Machine Vision and Signal Analysis, University of Oulu, Finland. Email: puneet.kumar@oulu.fi.} 
    \thanks{S. Malik, and B. Raman are with Indian Institute of Technology Roorkee, India. E-mail: sarthak\_m@mt.iitr.ac.in and bala@cs.iitr.ac.in}
    \thanks{X. Li is with the State Key Laboratory of Blockchain and Data Security, Zhejiang University, Hangzhou, China and the Center for Machine Vision and Signal Analysis, University of Oulu, Finland. E-mail: xiaobai.li@zju.edu.cn}
    \thanks{Manuscript received... ; revised...}
}
\markboth{Journal of \LaTeX\ Class Files,~Vol.~14, No.~8, August~2015}%
%\markboth{IEEE Transactions on Human–Machine Systems, Vol.~01, No.~1, Sep~2025}
{Shell \MakeLowercase{\textit{et al.}}: Bare Demo of IEEEtran.cls for IEEE Journals}

\maketitle

% =*=*=*=*=*=*=*=*=*=*=*=*=*=*=*=*=*=*=*=*=*=*=*=*=*=*=*=*=*=*
\begin{abstract} 
The ability to generate sentiment-controlled feedback in response to multimodal inputs comprising text and images addresses a critical gap in human-computer interaction. This capability allows systems to provide empathetic, accurate, and engaging responses, with useful applications in education, healthcare, marketing, and customer service. To this end, we have constructed a large-scale Controllable Multimodal Feedback Synthesis (CMFeed) dataset and proposed a controllable feedback synthesis system. The system features an encoder, decoder, and controllability block for textual and visual inputs. It extracts features using a transformer and a Faster R-CNN network, combining them to generate feedback. The CMFeed dataset includes images, texts, reactions to the posts, human comments with relevance scores, and reactions to these comments. These reactions train the model to produce feedback with specified sentiments, achieving a sentiment classification accuracy of 77.23\%, which is 18.82\% higher than the accuracy without controllability. Access to the CMFeed dataset and the system's code is available at \href{https://github.com/MIntelligence-Group/CMFeed}{github.com/MIntelligence-Group/CMFeed}. %The system includes a similarity module for evaluation of feedback relevance using rank-based metrics and an interpretability method to analyze textual and visual feature contributions during feedback generation.
\end{abstract}


%====================================
%-*-*-*-*-*- Introduction -*-*-*-*-*-
%====================================
\section{Introduction}\label{sec:intro}
The process of multimodal feedback synthesis involves generating responses to multimodal inputs in a way that mimics human spontaneous reactions \cite{kumar2023affective}. Controlling sentiments in feedback, a capability inherent to humans, remains a challenge for machines \cite{makiuchi2021multimodal}. The ability to control sentiments in feedback synthesis facilitates more empathetic responses in healthcare, accurate marketing insights, and engaging educational content while enabling systems to predict patients' mental states, assess product responses, analyze social behaviors, and gauge user engagement in advertisements \cite{gallo2020predicting, blikstein2016multimodal}. The controllability of these systems allows them to be customized for individual users, enhancing personalization \cite{muszynski2019recognizing}. Our goal is to advance human-centered, controllable interaction by explicitly regulating sentiment and exposing interpretable control signals, strengthening user understanding and trust in the loop.

The need for the Controllable Multimodal Feedback Synthesis (CMFeed) dataset arises from the requirement of a dataset containing human-generated feedback in addition to multimodal inputs. The CMFeed dataset has been created by crawling Facebook news articles and includes input images, texts, human comments, comments' metadata (such as likes, shares, reactions, and relevance scores), and sentiment labels. Unlike traditional sentiment-controlled text generation systems that do not utilize human comments, systems developed using the CMFeed dataset can be distinctively trained on human-generated comments to learn human-like spontaneity and contextual diversity. This enables the generation of `opinions' rather than just `knowledge' or `facts,' which the proposed task focuses on.  

In this work, a novel task of controllable feedback synthesis for input images and text has been defined, and a feedback synthesis system has been proposed to generate sentiment-controlled feedback. It includes two networks for textual and visual modalities, each with an encoder, a decoder, and a control layer. The encoders use a text transformer \cite{vaswani2017attention} and a Faster R-CNN model \cite{ren2015Faster} to extract features, which the decoder then combines for feedback generation. The control layer, positioned after the decoder, selectively activates or deactivates neurons to align with positive or negative sentiments as required. The system also incorporates a similarity module to ensure feedback aligns with the input context. The proposed system achieved a sentiment classification accuracy of 77.23\% on the CMFeed dataset, significantly higher than the baselines. The major contributions of this paper are summarized as follows:   

\begin{itemize}[leftmargin=*,itemsep=0pt]
\item A new dataset, CMFeed, has been constructed containing text, images, corresponding comments, number of likes, shares, and sentiment class. It enables training models to generate sentiment-controlled feedback from image–text inputs.
\item A feedback synthesis system capable of generating sentiment-controlled feedback has been developed. It extracts textual and visual features using transformer and Faster R-CNN models and combines them to generate feedback.
\item A novel controllability module has been proposed that selectively activates or deactivates neurons to align the feedback with the desired sentiment.
\item An interpretability technique, K-Average Additive exPlanation (KAAP) has been incorporated to analyze of textual and visual features' contribution during feedback generation.
\end{itemize}

% The remainder of the paper is organized as follows: the existing related works are surveyed in Section~\ref{sec:lr}. Section~\ref{sec:proposed} describes the dataset construction process and the proposed system. The experiments and results have been discussed in Section~\ref{sec:experiments} and Section~\ref{sec:results} respectively, and Section \ref{sec:conclusion} outlines the discussions, conclusions and future directions.

%=======================================
%-*-*-*-*-*- Related Works -*-*-*-*-*-*-
%=======================================
\section{Related Works}\label{sec:lr}
%This section explores the related works to the proposed task of controllable feedback synthesis.%, including multimodal summarization, visual question answering, dialogue generation, and sentiment-aware text generation.  

\subsection{Multimodal Summarization} 
Multimodal summarization combines inputs from various modalities to produce comprehensive summaries \cite{zhu2023topic}. Zhu et al. \cite{zhu2018msmo} focused on extractive summarization with a pointer generator; Li et al. \cite{li2020vmsmo} extended this to video with self-attention for frame selection. Other studies include Zhao et al. \cite{zhao2021audiovisual} on audio-visual summarization and Xie et al. \cite{xie2022multimodal} on visual aesthetics. Despite advances, multimodal summarization faces modality bias \cite{page2019multimodal, zhu2020multimodal}, is not explored for affect synthesis, and does not incorporate human comments in training. In contrast, our system is trained on human-generated comments alongside text and images to generate sentiment-controlled feedback; rather than condensing information, it produces feedback aligned with sentiments and context.

\subsection{Visual Question Answering}
Visual Question Answering (VQA) combines visual perception with interactive question answering \cite{antol2015vqa}. Chen et al. \cite{chen2023graph} explored controlled generation versus standard answering; Wang et al. \cite{wang2017fvqa} focused on fact-based control. Cascante et al. \cite{cascante2022simvqa} introduced synthetic data to expand scope, and Guo et al. \cite{guo2021universal} proposed a quaternion hypergraph network for multimodal video QA. Wu et al. \cite{wu2023memory} added memory-aware control in community QA, and Lehmann et al. \cite{lehmann2023language} used language models for knowledge graph QA, illustrating diverse applications and the shift toward vision and language pre-training. Unlike our task of sentiment-controlled feedback synthesis, which creates contextually relevant responses to multimodal stimuli (images and text), VQA focuses on accurate answers to questions.  

\subsection{Dialogue Generation} 
In dialogue generation, particularly visual dialogue (VisDial), computational models are designed to converse about visual content \cite{preece2017sherlock}. Kang et al. \cite{kang2019dual} developed a dual Attention Network that utilized the understanding of visual references in dialogues. Jiang et al. \cite{jiang2020dualvd} implemented a region-based graph attention network for enhanced image-related questioning. Contributions from Wang et al. \cite{wang2022unified} and Kang et al. \cite{kang2023dialog} enhanced dialogue generation through generative self-training. Liu et al. \cite{liu2023counterfactual} contributed to closed-loop reasoning and counterfactual visual dialogue for unbiased knowledge training. Despite these advancements, existing VisDial methods do not generate sentiment-controlled feedback. The proposed system stands apart as it synthesizes feedback that is contextually relevant and tailored to the sentiment of the conversation. 

\subsection{Sentiment-Aware and Sentiment-Controlled Generation}
In sentiment-aware conversational agents, Shi et al. \cite{shi2018sentiment} leverage user sentiments to guide dialogue actions; variational models predict emotional reactions to utterances for contextual guidance \cite{zhang2021predicting}; Firdaus et al. \cite{firdaus2020emosen} integrate sentiments for sentiment-controlled personalized responses; Hu et al. \cite{hu2022acoustically} design a speech sentiment-aware agent. Controllability has also extended to other modalities, e.g., Music ControlNet for melody/rhythm \cite{wu2024music} and glyph-conditioned control for text-to-image precision \cite{yang2024glyphcontrol}. However, sentiment-controlled text generation does not use human comments. By contrast, our system captures wider sentiment context from multimodal (text and images) inputs and is trained on human-generated comments to learn contextual diversity. Our work generates `opinions’ rather than `knowledge’ or `facts’; unlike facts, opinions can be controlled similarly to humans. This area has not been addressed by earlier tasks that our work advances.

\begin{table*}[]
\centering
\caption{Summary of the existing related datasets. Here, `HC': Human Comments, `CM': Comments' Metadata, `V': Visual, `A': Audio, `T': Textual, `MMSum': Multimodal Summarization, `VQA': Visual Question Answering, `VisDial': Visual Dialogue Generation, `SATextGen': Sentiment-Aware Text Generation, and `N/A': Not Applicable, `-': Unavailable.}
\label{tab:datasets}
\resizebox{.88\textwidth}{!}{
\begin{tabular}{ccccccccc}
\toprule
\textbf{Area} & \textbf{Dataset} & \textbf{Year} & \textbf{Dataset Size} & \textbf{No. of Subjects} & \textbf{Modalities} & \textbf{HC} & \textbf{CM} \\  %& \textbf{Description and Focus Area} 
\midrule
\multirow{4}{*}{\rotatebox[origin=c]{90}{\makebox[1.1cm][c]{MMSum}}} 
& VMSMO \cite{li2020vmsmo}
%& Video and text summarization for news articles use multimodal attention. 
& 2020 & 184920 documents & 70 participants & V, T & \ding{51} & \ding{55} \\
 
& MSMO \cite{zhu2018msmo}
%& Extractive and abstractive summarization with multimedia outputs, focusing on content compression. 
& 2018 & 314581 documents & 10 students & V, T & \ding{51} & \ding{55} \\

& MVSA-Single \cite{niu2016sentiment}
%& Sentiment analysis in tweets with single images, used for understanding public sentiments. 
& 2016 & 4869 tweets & N/A & V, T & \ding{55} & \ding{55} \\

& MVSA-Multiple \cite{niu2016sentiment}
%& Extension of MVSA-Single with multiple images per tweet. 
& 2016 & 4869 tweets & N/A & V, T & \ding{55} & \ding{55} \\
\midrule

\multirow{4}{*}{\rotatebox[origin=c]{90}{\makebox[0.1cm][c]{VQA}}} 
& DocVQA \cite{mathew2020document}
%& Task of question answering on document images, suitable for OCR and document understanding. 
& 2020 & 50K QA pairs & - & T & \ding{55} & \ding{55}  \\

& OK VQA \cite{marino2019ok} 
%& Knowledge-based visual question answering that requires external information beyond the image content. 
& 2019 &  150K QA pairs &  5 MTurk workers & V, T & \ding{55} & \ding{55}  \\

& VQA \cite{goyal2017making} 
%& Standard dataset for image-based question answering focuses on object recognition. 
& 2017 & 1.1M QA pairs & 215 MTurk workers & V, T & \ding{51}  & \ding{55} \\

& VideoQA \cite{xu2017video} 
%& Question answering based on video content focuses on dynamic scene understanding. 
& 2017 & 243K QA pairs  & -  & V, T & \ding{55} & \ding{55}  \\
\midrule

\multirow{4}{*}{\rotatebox[origin=c]{90}{\makebox[1.1cm][c]{VisDial}}} 
& InfoVisDial \cite{wen2023infovisdial}
%& Visual dialogue focusing on informative responses with relevance to the visual content. 
& 2023 & 79535 dialogues & 35 annotators & V, T & \ding{55} &\ding{55} \\

& CLEVR-Dialog \cite{kottur2019clevr}
%& A synthetic dataset that facilitates visual reasoning in dialogue systems. 
& 2019 & 4.25M dialogues & - & V, T & \ding{55} &\ding{55} \\

& VisDial \cite{visdial} 
%& Engages models in a dialogue about visual content, focusing on conversational understanding. 
& 2017 & 1.23M QA pairs & 200 annotators & V, T & \ding{51}  & \ding{55} \\

& Visual Madlibs \cite{yu2015visual}
%& Fill-in-the-blank questions about images, tests specific understanding of visual scenes. 
& 2015 & 397675 dialogues & 1 quality checker & V, T &\ding{55} &\ding{55} \\  
\midrule

\multirow{4}{*}{\rotatebox[origin=c]{90}{\makebox[1.1cm][c]{SATextGen}}} 
& SEPRG \cite{firdaus2021seprg}
%& Personalized response generation with emotion control, focuses on user-specific dialogue. 
& 2021 & 64356 conversations & 500 samplers & T & \ding{51}  & \ding{55}\\

& EMOTyDA \cite{saha2020towards}
%& Emotion detection in dialogue systems provide emotion annotations for responses. 
& 2020 &  19365 videos & 10 participants & V, T & \ding{51}  & \ding{55}\\

& ESTC \cite{zhou2018emotional}
%&Emotional short-text conversations designed to handle dialogues with emotional context. 
& 2018 & 4308211 conversations & - & T & \ding{51}  & \ding{55}\\

& STC \cite{shang2015neural}
%& Short-text conversations aimed at building dialogue systems include casual everyday topics. 
& 2015 & 4.4M conversations & - & T & \ding{51}  & \ding{55}\\
\bottomrule
\end{tabular}
\vspace{-.5in}
}
\end{table*}

Existing datasets are summarized in Table \ref{tab:datasets}. While some include human-generated summaries, none provide metadata such as relevance, reactions, or sentiment labels for feedback synthesis systems. The lack of human comments and metadata limits sentiment-controlled feedback synthesis \cite{kumar2023affective, wu2019response}. CMFeed fills this gap by enabling models to generate opinions that mimic human conversational dynamics, supporting context-aware multimodal interactions. Unlike large language models (LLMs) such as GPT \cite{radford2018improving}, which focus on summarization and question answering, CMFeed interprets multimodal inputs to produce sentiment-controlled feedback. LLMs are prone to errors and hallucinations \cite{farquhar2024detecting, huang2023survey}, are autoregressive, and obscure modality influence. Our interpretability mechanism shows how metadata and multimodal features shape responses and sentiment, and these notions of controllability and interpretability can inform future LLM and VisDial research.

%=======================================
%-*-*-*-*-*- Proposed Method -*-*-*-*-*-
%=======================================
\section{Proposed Methodology}\label{sec:proposed}
\subsection{Dataset Construction}\label{sec:dataset}
\subsubsection{Data Collection}
The CMFeed dataset has been compiled by crawling news articles from Sky News, NYDaily, FoxNews, and BBC News through Facebook posts. The data collection process utilized the NLTK \cite{nltk} \& newspaper3k \cite{newspaper3k} libraries and it was conducted in compliance with Facebook's terms and conditions \cite{facebookterms}, ensuring legal and ethical standards. \vspace{.05in}

\noindent \textit{Choice of Facebook for Dataset Construction}: 
Facebook was chosen for data collection due to its unique provision of metadata, i.e., news article links, post shares, post reactions, comment likes, comment rank, comment reaction rank, and relevance scores are unavailable on other platforms. With $3.07$ billion monthly users, it is the most used social media platform, compared to Twitter’s 550 million and Reddit’s 500 million \cite{wiki_socialpf}. Popularity holds across age groups, with at least 58\% usage among ages 18–29, 30–49, 50–64, and 65+, versus 6\% for Twitter and 3\% for Reddit \cite{fb_vs}. This trend is consistent across gender, race/ethnicity, income, education, community, and political affiliation \cite{other_trends}. The male-to-female user ratio is 56.3\% to 43.7\% on Facebook, compared to Twitter’s 66.72\% to 23.28\%; such data are not reported by Reddit \cite{fbm2f}.\vspace{.05in} %Together, these characteristics make Facebook an ideal platform for collecting a diverse, representative dataset.

\noindent \textit{Choice of News Handles}: Four news handles (BBC News, Sky News, Fox News, and NY Daily News) were selected to ensure diversity and comprehensive regional coverage. These news outlets were chosen for their distinct editorial perspectives and regional focuses: global coverage and centrist views are known from BBC News, targeted content and center/right leanings in the UK, EU, and US are offered by Sky News, right-leaning content in the US is recognized from Fox News, and left-leaning coverage in New York is provided by NY Daily News. A broad spectrum of political discourse and audience engagement is ensured by this selection. 

\subsubsection{Preprocessing} \label{sec:prepro} 
The CMFeed dataset includes multiple images per sample, corresponding news text, post likes and shares, and human comments with reactions and shares. Comments are sorted by Facebook’s `most-relevant’ criterion, prioritizing the highest likes and shares. Preprocessing converts emoticons to words with Demoji \cite{demoji}, removes blank comments, expands contractions, eliminates special and accented characters, purges punctuation and numbers, removes stop-words, and lowercases text. It reduces noise while retaining key sentiment cues (core sentiments are conveyed mainly through words rather than punctuation or numbers). As per Facebook’s ethical protocols, we used manual scraping and selected $1000$ posts from each of four news handles. Initially $4000$ posts were collected; after preprocessing, $3646$ remained, with all associated comments processed, yielding $61734$ comments. As summarized in Table \ref{tab:data1}, each news post received on average $65.1$ likes, comments averaged $10.5$ likes, news text averaged $655$ words, and each post contained on average $3.7$ images.\vspace{-.05in}

\begin{table}[!h]
  \centering
  \caption{CMFeed dataset's overview.}
  \label{tab:data1}
  \resizebox{0.34\textwidth}{!}{% Resize table to fit within the wrapping text width
    \begin{tabular}{@{}lc@{}}
        \toprule
        \textbf{Parameter}               & \textbf{Value} \\ \midrule
        No. of news posts                & 3646          \\
        No. of total data samples        & 61734         \\
        No. of samples after filtering   & 57222         \\
        Avg. no. of likes per post       & 65.1           \\
        Avg. no. of likes per comment    & 10.5           \\
        Avg. length of news text         & 655 words      \\
        Avg. no. of images per post      & 3.7            \\ \bottomrule
    \end{tabular}
  }
\end{table}

\begin{table*}[]
\centering
\caption{Representative samples from the CMFeed dataset. Here, `PLikes’ and `CLikes’ are likes on the post and the comment, `Shares’ is the post’s share count and `C’ is the comment's sentiment class (1 positive, 0 negative).}
\label{tab:datasamples}
\resizebox{1\textwidth}{!}{
\begin{tabular}{m{1.7cm}m{5.7cm}C{3.2cm}C{.6cm}C{.6cm}m{1.5cm}C{.5cm}C{.5cm}}
\hline
\textbf{Title} & \textbf{Text} & \textbf{Images} & \textbf{PLikes} & \textbf{Shares} & \textbf{Comment} & \textbf{CLikes} & \textbf{C} \\ \hline

There are costs of managing beavers, but the benefits outweigh those costs. 
& Beaver dams in east Devon create area of wetland amid drought, The dams have created a wetland despite the dry weather. A network of dams built by beavers in Devon has helped to maintain an area of wetland despite a drought in the South West. There are a number of beavers ...
&\includegraphics[width=3.2cm]{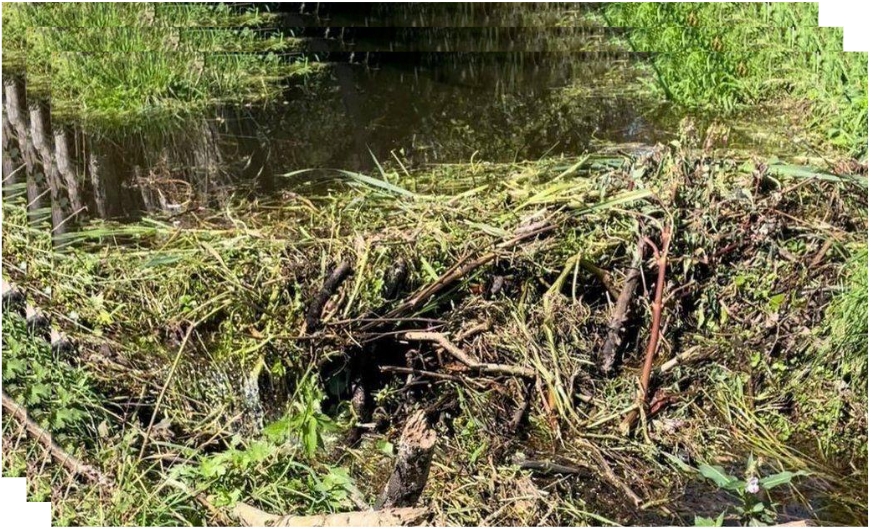} 
& 2887 
& 165 
& Benefits outweigh the costs because beavers are ecosystem engineers! 
& 47
& 1 \\ \hline 

A national emergency has been declared. 
& Pakistan floods: Monsoons bring misery to millions in PakistanBy Pumza Fihlani in Sukkur, Pakistan and Frances Mao in Singapore. Millions of people have been affected by floods in Pakistan, hundreds have been killed, and the government has declared a national emergency ...
& \includegraphics[width=3.2cm]{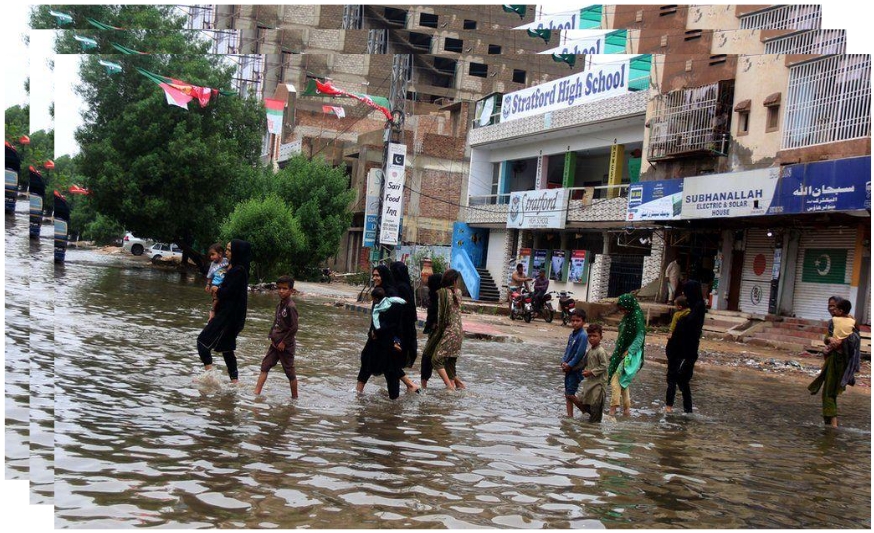} 
& 2005 
& 126 
& Circumstances are really miserable after monster floodings. 
& 26
& 0 \\ \hline

Celebrating his birthday, John Tinniswood said moderation in everything.
& Moderation is the key to life, GB's oldest man says on 110th birthday, Mr Tinniswood was joined by family and friends to celebrate his big day. Britain's oldest man has celebrated his 110th birthday by declaring ``moderation in everything and all things'' as the secret ... 
& \includegraphics[width=3.2cm]{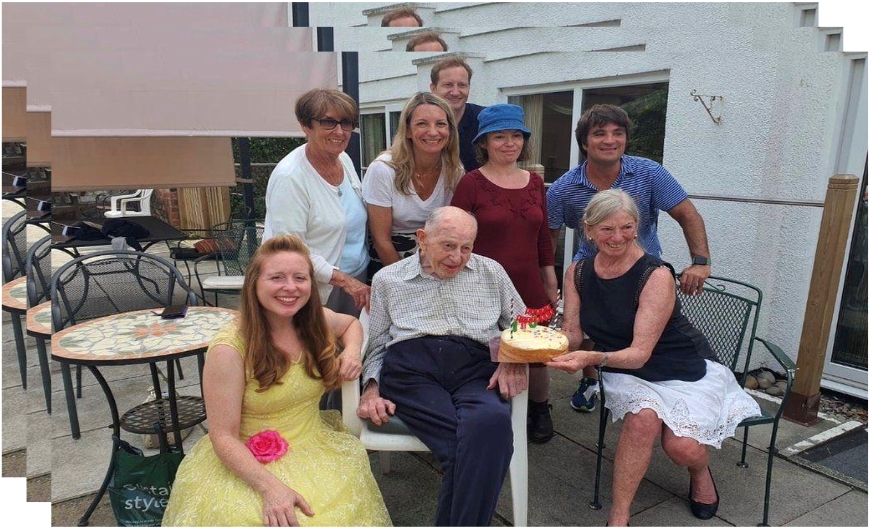} 
& 12000 
& 407 
& Congratulations on a well lived life and 110.
& 31
& 1 \\ \hline

% If it leaks then it's easy to clean. 
% & Bin strikes: The people using baths and hiring skips to store rubbish, Helen Sikora has been keeping rubbish bags in her bath so she can easily clean up any leaks. Edinburgh residents have told how they have hired skips and used bathtubs to store rubbish, as waste piles up ... 
% & \includegraphics[width=3.2cm]{fig_4.jpg} 
% & 464 
% & 29 
% & Dump it on streets if they do not bother to collect it.
% & 29
% & 0 \\ \hline

\multicolumn{8}{c}{\Large {...}} \\ \hline
\end{tabular}
}  \vspace{-.05in}
\end{table*}

\subsubsection{Annotation Strategy} To determine the ground-truth sentiment labels for the comments, we obtained sentiment scores using four pre-trained models: FLAIR Library \cite{akbik2019flair}, SentimentR \cite{sentimentr}, DistilBERT \cite{sanh2019distilbert}, and RoBERTa \cite{liu2019roberta}. Each of these models has unique capabilities: FLAIR specializes in capturing contextual variations in text using a neural network approach. SentimentR is designed to analyze textual sentiments by evaluating linguistic cues within the text. DistilBERT and RoBERTa are both transformer-based models optimized for understanding the nuances of language through self-attention mechanisms. DistilBERT offers a lighter, faster variant of BERT that retains most of its predictive power, whereas RoBERTa is trained on an even larger corpus with more robust fine-tuning, enhancing its ability to discern complex sentiment patterns.

We adopted a majority voting strategy for annotation, as in IEMOCAP \cite{busso2008iemocap}, retaining samples that received the same sentiment class from at least three of four models; the rest were excluded and marked `XX.' The models output sentiment values of $-1$ or $1$ with a confidence score. We computed a score by multiplying sentiment and confidence, normalized it to $[0,1]$ with $0$ negative and $1$ positive, and averaged the normalized scores across models to obtain final labels. To ensure robustness, we used a safety margin from $0.49$ to $0.51$ and marked labels in this range as `XX.' $57222$ of $61734$ samples met the criteria, providing high-confidence ground-truth labels.

The sentiment prediction was conducted on comments to capture and analyze direct human emotional reactions, independent of the original text or imagery. This approach aims to understand user reactions, which are direct, spontaneous, and personal, thus providing insights into user sentiments. The objective is to enable the feedback synthesis system to emulate this human-like directness and spontaneity. To ensure the accuracy of the constructed ground-truth sentiment labels, a human evaluation was conducted. $50$ evaluators ($25$ males and $25$ females, average age $30 \pm 2.73$ years) assessed the sentiment of $50$ randomly selected pairs of input image and text. The results showed that $90.88$\% of evaluators (standard deviation $7.59$\%) agreed on the consistency between the expressed sentiment and the assigned sentiment label. 

The CMFeed dataset's samples have been depicted in Table \ref{tab:datasamples}. Multiple images and comments correspond to each news post, enabling the feedback synthesis model to learn the comments' contextual diversity and relevance to the input. The CMFeed dataset and the corresponding code can be accessed at \href{https://zenodo.org/records/11409612}{zenodo.org/records/11409612} and \href{https://github.com/MIntelligence-Group/CMFeed/}{github.com/MIntelligence-Group/CMFeed/}, respectively. More details about the CMFeed dataset are provided in Supplementary Material Sec.~S2. %along with detailed description of its metadata, collection process, intended usage, licensing, and handling. T

\subsection{Task Formulation} 
Given an environment \( E = \{T, I_1, I_2, \ldots, I_n\} \), where \( T \) denotes input text and \( I_1, I_2, \ldots, I_n \) denote \( n \) input images. Each image comprises \( m \) objects \( o_1, o_2, \ldots, o_m \), while the text \( T \) is made up of a dictionary of \( k \) words \( w_1, w_2, \ldots, w_k \). The task is to generate a feedback towards environment \( E \) in a sentiment-controlled manner where `sentiment-controlled' implies the feedback aligns with a specified sentiment $S$, with $S$ being either $0$ or $1$ for a negative or positive sentiment. 

\subsection{Proposed Feedback Synthesis System}\label{sec:prop}
\begin{figure*}
\centering
\includegraphics[width=1\textwidth]{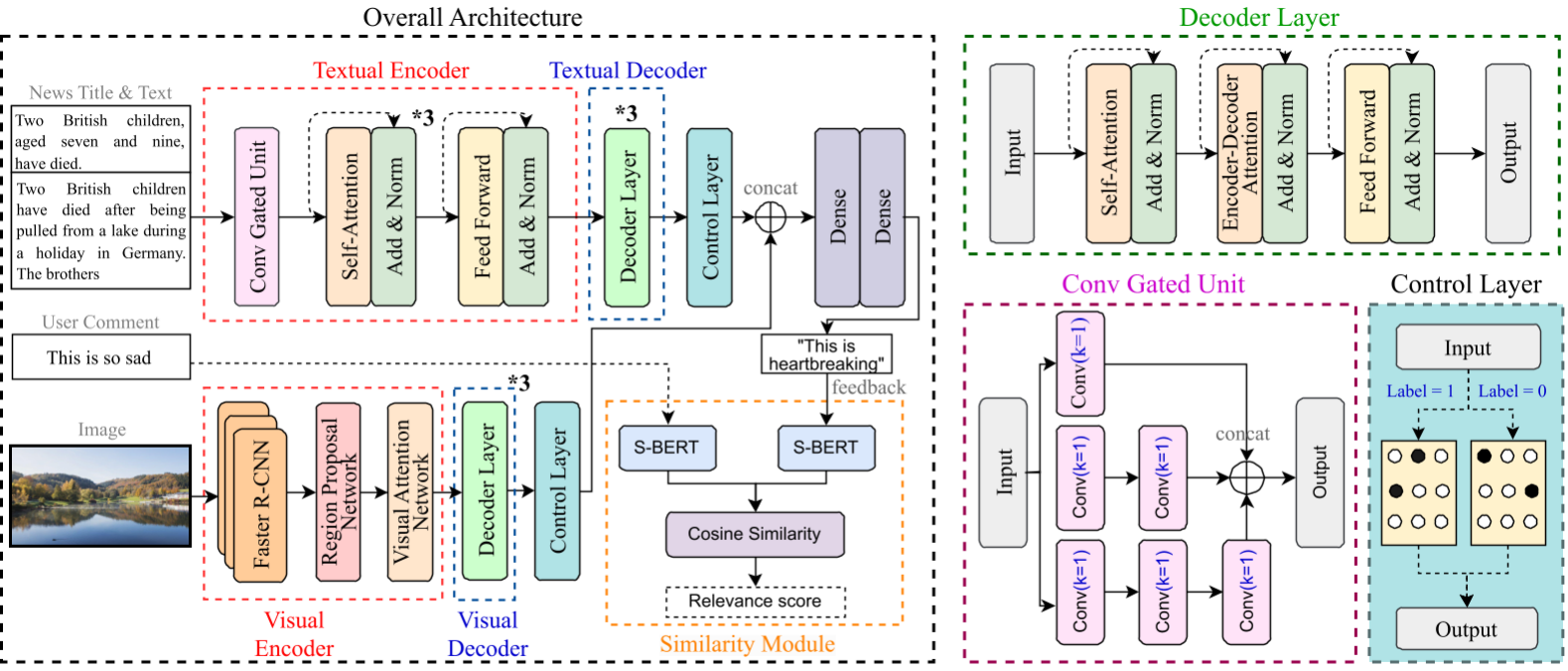}\vspace{-.05in}
\caption{Proposed system's architecture with encoder, decoder and controllability blocks for textual and visual data. Decoder, convolution-gated unit, control layer, and similarity modules appear as subblocks; on/off neurons as black/white circles.\vspace{-.05in}} 
\label{fig:methodology}	
\end{figure*}

The proposed system (Fig. \ref{fig:methodology}) includes two networks for textual and visual data, each with an encoder, a decoder, and a control layer. The system generates context-aware feedback matching the desired sentiment whose alignment with human comments is checked by the similarity module. 

\subsubsection{Textual Encoder}%\ref{https://arxiv.org/pdf/1805.03989.pdf}
The textual encoder utilizes a transformer model \cite{vaswani2017attention}, featuring global encoding and textual attention mechanisms. Global encoding is enhanced by a convolution-gated unit to improve textual representation, reduce repetition, and maintain semantic relevance. The textual attention component includes multi-headed self-attention, comprising a self-attention layer and a feed-forward layer. Positional embedding captures token positioning, and normalization finalizes the process to produce the textual context vector $z_t^*$. The feed-forward network (FFN) includes input and output layers with a dimension of $512$ and a hidden layer of $2048$. The FFN's output for a specific input $x$ is defined in Eq.\ref{eq:tenc}.\vspace{-.1in}

{%\fontsize{9}{11}\selectfont
\begin{equation}
\label{eq:tenc}
FFN(x) = max(0, xW_1 + b_1)W_2 + b_2 \
\end{equation}
}\vspace{-.2in}

\noindent where, $b_1$, $b_2$, $W_1$, and $W_2$ represent the bias terms and weight matrices, respectively. In the self-attention mechanism, the query, key, and value weight matrices are initially randomized in the encoder and updated during training.

\subsubsection{Visual Encoder}
The top three images from each sample are input to a visual encoder, using blank images when fewer are available; their features are concatenated to form the visual context vector $z_i^*$. We employ a pre-trained Faster R-CNN \cite{ren2015Faster}: CNN layers yield feature maps that the RPN converts to anchor boxes with binary scores based on IoU \cite{rosebrock2016intersection}, which are then classified and regressed into $1601$ classes, whose features are combined into a global vector. Faster R-CNN is selected for efficiency and precision in detecting small and varied objects in non-real-time settings. Using the top three images balances retaining essential visual content and minimizing blanks, since most posts have at least three images.\vspace{-.15in} 

{%\fontsize{8}{10}\selectfont
\begin{equation}
\label{eq:venc}
\text{\shortstack[l]{Objectness\\Score}}
\;=\;
\begin{cases}
  \text{Positive;} \quad \mathrm{IoU} > 0.7 \\
  \text{Negative;} \quad \mathrm{IoU} < 0.3 \\
  \text{No score;} \quad 0.3 < \mathrm{IoU} < 0.7
\end{cases}
\end{equation}
}\vspace{-.1in}

\subsubsection{Attention}
The attention mechanism in both the encoder and decoder operates on three vectors: $Q$ (query), $K$ (key), and $V$ (value). The output of the self-attention layer, denoted as $z_i$, is computed by multiplying the $i^{th}$ input vector of the encoder with the respective weight matrices $W(Q)$, $W(K)$, and $W(V)$. This computation yields the attention head matrix $z$, as detailed in Eq. \ref{eq:visatn}, whose dimensionality is equivalent to the length of the input sequence.\vspace{-.2in}

{%\fontsize{9}{11}\selectfont
\begin{eqnarray}
\label{eq:visatn}
\begin{split}
&\ \ \ z = Attention(Q, K, V ) = softmax(\frac{Q.K^T}{\sqrt{d_k}})V\
\end{split}
\end{eqnarray}
}\vspace{-.1in}

\noindent where $Q$, $K$, and $V$ are matrices containing all queries, keys, and values, with $d_k$ as scaling factor and $K^T$ transpose of $K$. To achieve a comprehensive subspace representation, the mechanism computes multiple attention heads using distinct Query, Key, and Value projections. The queries, keys, and values undergo projection $head$ times, yielding heads $h_1, h_2, ..., h_{head}$, where $head$ is the total number of heads. The heads are then concatenated and multiplied by the weight matrix $W$, producing the intermediate output vector $z'$, as in Eq.~\ref{eq:att}.\vspace{-.15in}

{%\fontsize{9}{11}\selectfont
\begin{eqnarray}
\label{eq:att}
\begin{split}
&h_i = Attention(QW^{Q}{i}, KW^{K}{i}, VW^{V}{i}) \\
&z' = Concat(h_1, h_2 ..., h_{head})W^O\
\end{split}
\end{eqnarray}
}\vspace{-.05in}

\noindent where $W^{Q}{i}$, $W^{K}{i}$, $W^{V}{i}$, and $W^{O}{i}$ are the respective projections of queries, keys, values, and output of corresponding heads. The final context vector $z^*$ is then derived by passing this intermediate output through the feed-forward layer.

\subsubsection{Decoder}
The textual and visual decoders share a similar structure: two blocks (self-attention and encoder–decoder attention) with positional encoding and normalization. The textual decoder processes $z_t$, the visual decoder handles $z_i$; both also receive the ground-truth comment. The self-attention layer uses future position masking to attend only to previous outputs. The encoder–decoder attention uses $z_t$ or $z_i$ as keys and values. Late fusion via concatenation preserves distinct image features, and a gated convolution unit is added for textual features to reduce repetition.

\subsubsection{Control Layer} \label{sec:Controllability}
The control layer, positioned after the decoder and before feedback generation, introduces perturbations to ensure feedback aligns with desired sentiments. It utilizes two masks, one each for positive and negative sentiments, altering the input vector via element-wise multiplication as per Eq. \ref{eq:control1}. This layer functions like a modified dropout layer, selectively activating or deactivating neurons to tune sentiment in the feedback, ensuring it matches the targeted tone. From a human–machine systems perspective, providing interpretable control signals improves user understanding and trust in the loop. Our mask-based control layer follows this principle by including an inspectable intervention on the decoder state \cite{ji2022improving}.\vspace{-.1in}

{%\fontsize{9}{11}\selectfont
\begin{equation}
\label{eq:control1}
O =
\begin{cases}
    mask_1*I; \hspace{.35in} Sentiment = 0   \\ 
    mask_2*I; \hspace{.35in}  Sentiment = 1  \\
\end{cases}
\end{equation}
}\vspace{-.1in}

\noindent where $O$ and $I$ denote the output and input vectors, respectively. Each mask blocks $x\%$ neurons, targeting different neuron sets; consequently, $(100-2x)\%$ of neurons are trained on both sentiments, while $x\%$ are specialized, with $x=10\%$. This directs the feedback toward the desired sentiment. At inference, to generate sentiment-specific feedback, neurons trained for the contrasting sentiment are deactivated; for instance, to produce positive feedback, neurons associated with negative sentiment are turned off, and vice versa. The output sentiment is controlled independently of the input text’s sentiment, steering the generated sentence’s sentiment.

\subsubsection{Similarity Module}\label{sec:similarity} 
The similarity module quantitatively assesses semantic similarity between generated feedbacks and human comments using a pre-trained Sentence-BERT (SBERT) model \cite{reimers2019sentence}. It transforms comments and feedbacks into vectors in an $n$-dimensional embedding space (where $n$ is the model’s embedding size), with each dimension representing a distinct linguistic feature capturing sentence characteristics. Cosine similarity is then computed between the feedback and comment vectors. This score captures the orientation of the sentence vectors, reflecting semantic similarity and providing a relevance measure between feedbacks and comments.

\subsubsection{Interpretability}\label{sec:interpretability}
\begin{figure}[]
  \centering
  \includegraphics[width=.5\textwidth]{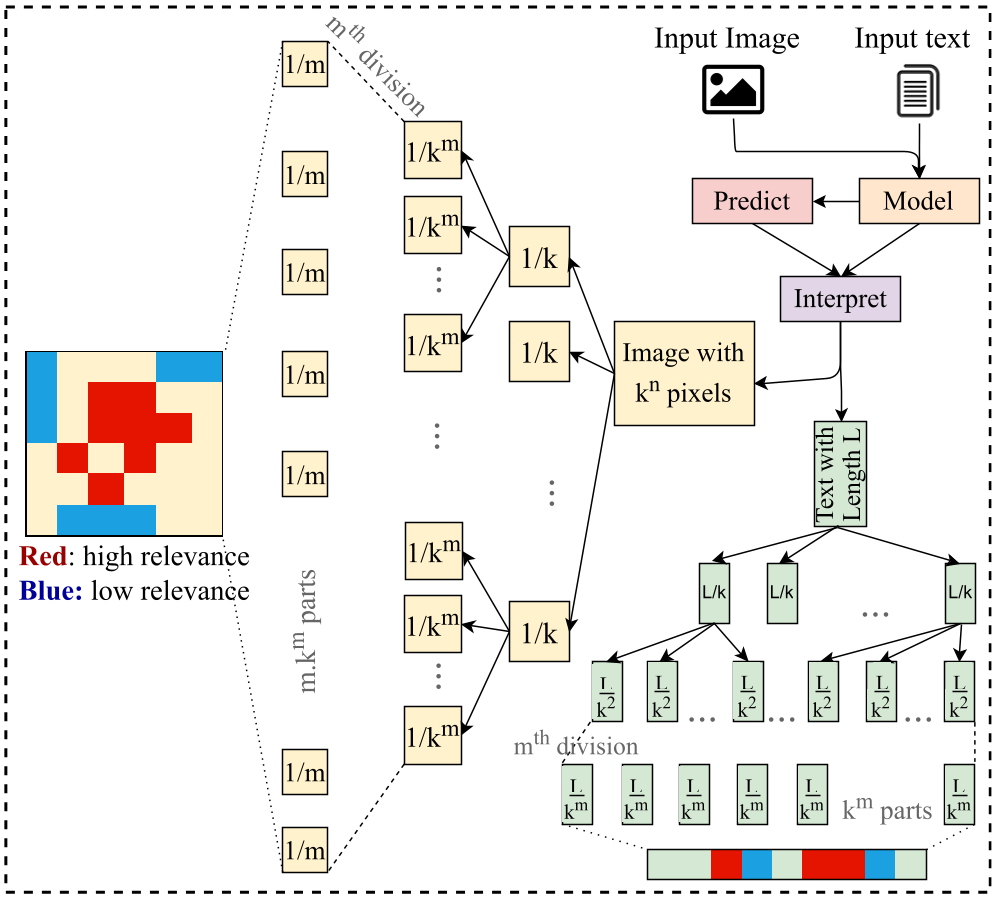}  
  \vspace{-.2in}
  \caption{Depiction of the proposed interpretability technique. Here \(k_i\) and \(k_t\) are number of partitions for image and text, \(w_i\) is the image's width and \(L_t\) is the text feature vector's length.\vspace{-.1in}}
  \label{fig:inter} 
\end{figure}

This section proposes an interpretability technique using the K-Average Additive exPlanation (KAAP) method \cite{kumar2025vistanet} to analyze the contribution of textual and visual features toward feedback generation. It is based on an approximation of Shapley values \cite{SHAPley1953value}. As depicted in Fig. \ref{fig:inter}, it assesses the influence of each visual and textual feature on the sentiment portrayed by the generated feedback. We hypothesize that varying sentiments produced from identical inputs (text + images) should reflect in differential feature importance, with key features differing for negative versus positive sentiments. When identical inputs portray varied sentiments, the model should adjust its focus across image and text segments, thereby validating our controllability hypothesis.

% \begin{wrapfigure}{r}{0.8\textwidth} % "r" right; "l" left
% \end{wrapfigure}

\vspace{.05in}\noindent\textit{SHAP Values Computation}: 
The SHAP values for the features denote their contribution to the model's prediction. For a model \( f \), the SHAP value for feature \( i \) is defined as per Eq. \ref{eq:shap}. \vspace{-.1in}

\begin{equation}
\label{eq:shap}
\mathscr{S}_{i}(f) = \sum_{S \subseteq F \setminus \{i\}} \frac{|S|!(|F| - |S| - 1)!}{|F|!} [f(S \cup \{i\}) - f(S)]
\end{equation} 
\vspace{-.07in}

\noindent where \( F \) denotes the complete feature set, \( S \) a subset excluding \( i \), and \( f(S) \) the model's prediction using features in \( S \).

The computation of SHAP values requires exponential time theoretically which is approximated by dividing the input into $k$ parts as illustrated in Eq. \ref{eq:k}. For each modality, the input is repeatedly divided into $k$ segments, determining each segment's impact on model predictions. A feature vector \(X\) with \(n\) features is segmented into \(k\) parts. \vspace{-.25in}

\begin{equation}
\label{eq:k}
X = [X_1, X_2, ..., X_k], \text{where } X_i \subseteq X \text{and} \bigcup_{i=1}^{k} X_i=X
\end{equation} 
\vspace{-.1in}

For simplicity with $k = 2$, the fundamental computation of SHAP values is denoted in Eq. \ref{eq:shap_basic}. It is extended for other values of $k$. The optimal values of \(k_{\text{img}}\) and \(k_{\text{txt}}\), representing the number of segments to divide the image and text into, have been determined experimentally.\vspace{-.1in}

\begin{equation}
\label{eq:shap_basic}
\mathscr{S}_{\{f_1\}} + \mathscr{S}_{\{f_2\}} = \mathscr{S}_{\{f_1, f_2\}} - \mathscr{S}_{\{\text{null}\}}
\end{equation}
\vspace{-.15in}

\noindent\textit{K-Average Additive exPlanation (KAAP)}:
The KAAP value for feature \( i \) is calculated by averaging the SHAP values across the \( k \) divisions of the feature vector using Eq. \ref{eq:kaap}.\vspace{-.1in}

\begin{equation}
\label{eq:kaap}
\text{KAAP}_i = \frac{1}{k} \sum_{j=1}^{k} \mathscr{S}_{i}(X_j)
\end{equation}
\vspace{-.1in}

The KAAP values directly indicate the significant image features for predictions. For input image \( X^{img} \) of dimensions \(128 \times 128\), the KAAP values for a given \(k\) are computed by segmenting the input along both axes. For text data \( X^{txt} \), we derive the feature vector and divide it into \( k \) segments. Text division considers each word as a feature, acknowledging that sentiments are conveyed by words, not individual letters.

%-*-*-*-*-*- Experiments-*-*-*-*-*-
\section{Experimental Results}\label{sec:experiments}	
\subsection{Training Strategy and Parameter Tuning}
The proposed models were trained for $60$ epochs on an Nvidia V$100$ GPU using $5$-fold cross-validation with an $80$–$20$ train–test split. General parameters included a batch size $16$, learning rate $0.001$, Adam optimizer, cross-entropy loss, and ReLU activation. For the transformer, encoder and decoder embeddings were $100$, hidden units $128$ (encoder and decoder), dropout $0.1$ (encoder and decoder), with $3$ layers and $8$ attention heads for both encoder and decoder; the metric was accuracy. For Faster R-CNN, training used $18$ epochs, $18$ proposals, $1601$ anchor-box classes, the AdaDelta optimizer, and mean Average Precision as the metric.

\subsection{Evaluation Metrics}\label{sec:metrics}
Feedback synthesis is a one-to-many task, i.e., many feedbacks can be generated for one pair of image-text input. Hence, computing the accuracy of generated feedbacks is not feasible. Instead, we evaluate the generated feedbacks against the ground-truth comments using the metrics to evaluate semantic relevance and their ranks. For semantic relevance, we use BLEU \cite{papineni2002bleu}, CIDEr \cite{vedantam2015cider}, ROUGE \cite{lin2004rouge}, SPICE \cite{anderson2016spice}, and METEOR \cite{lavie2009meteor}. BLEU is precision-based, ROUGE is recall-based, METEOR combines both, CIDEr is consensus-based, and SPICE evaluates semantic structures or n-grams. Higher values of these metrics denote more semantic similarity between the feedback and ground-truth comment. For ranking-based evaluation, we use `Mean Reciprocal Rank' and `Recall@k'. \vspace{.07in}

%\begin{itemize}[leftmargin=*]
%\item 
\noindent {Mean Reciprocal Rank} \cite{craswell2009mean}:
For Mean Reciprocal Rank (MRR), each generated feedback is compared with all ground-truth comments; if the most similar comment is ranked $k$, the $j^{th}$ feedback has $rank_j = k$ (the $k^{th}$ comment when sorted by relevance). MRR is the average reciprocal rank over all samples: $MRR = \frac{1}{n}\sum_{j=1}^{n}\frac{1}{rank_j}$, where $n$ is the number of generated feedback samples and $rank_j$ is the rank of the $j^{th}$ feedback. \vspace{.07in}

%\item  
\noindent {Recall@k} \cite{runeson2007detection}: Recall@k counts the number of data samples matching any top-$k$ relevant samples. For generated feedback, we count feedbacks similar to any of the top-$k$ comments sorted by relevance. If a feedback’s rank is in the top-$k$, it receives a score of $1$, else $0$. Recall@k is then $\sum_{i=1}^{n} Recall@k_i$, where $Recall@k_i$ and $rank_i$ are the $i^{th}$ feedback’s Recall@k and rank. Sentiments of generated feedbacks are also computed, and sentiment classification accuracy is analyzed along with the `\textit{Control Accuracy}', the difference between accuracies of controlled and uncontrolled feedbacks. %Recall@k_i = 1\ \text{if}\ rank_i \in [1, \dots, k] To find if the generated feedback is similar to any comment, the rank of that feedback as calculated in Eq. \ref{eq:mrr} is used. Finally, the Recall@k can be formulated according to Eq. \ref{eq:recall}. 
%\end{itemize}

\subsection{Models}
%The ablation studies discussed in Section~\ref{sec:ablation} have informed the design of both the baseline and proposed model architectures. 
The following models' architectures are determined through ablation studies (Section \ref{sec:ablation}). Each includes the Controllability module, with the remaining architecture as follows

\begin{itemize}[leftmargin=*, itemsep = 1pt]
  \item \textit{Baseline 1} utilizes Gated Recurrent Units (GRU) as textual and VGG network as visual encoders. An early fusion method is applied to integrate visual and textual modalities. 
  \item \textit{Baseline 2} uses a late fusion approach for combining the visual and textual data while maintaining GRU for textual encoding and VGG for visual encoding.
  \item \textit{Baseline 3} implements a combination of a Transformer and a gated convolutional unit for textual encoding. It uses Faster RCNN with visual attention mechanism for visual encoding and a late fusion strategy with averaging.
  \item \textit{Baseline 4} replaces the textual encoder with GPT-2 \cite{gpt2} and continues to use Faster RCNN with visual attention for visual data encoding and late fusion with averaging. It has been empirically observed that GPT-$2$ based model generated good feedbacks only for textual input; however, it did not generate good feedbacks for multimodal input.
  \item \textit{Proposed System} incorporates Transformer as the textual encoder and Faster RCNN as visual encoder and it uses concatenation along with late fusion. 
\end{itemize}

%-*-*-*-*-*- Results-*-*-*-*-*-
\subsection{Results}\label{sec:results}
%The results of the proposed benchmark feedback synthesis system are presented as follows.

\subsubsection{Semantic Relevance Evaluation}
The generated feedbacks' semantic relevance with human comments has been evaluated. The feedbacks are generated to reflect the same sentiment class as reflected by the corresponding comments and then the feedbacks are evaluated using the BLEU, CIDEr, ROUGE, SPICE, and METEOR metrics. As depicted in Table~\ref{tab:quantre}, the proposed model has obtained the best values for these metrics in most cases. 

\begin{table}[!h] 
\centering
{%\fontsize{9}{11}\selectfont
    \caption{Semantic Relevance Evaluation.}
    \label{tab:quantre}
    \resizebox{.48\textwidth}{!}{
        \begin{tabular}{@{}l|ccccc@{}}
            \toprule
            \textbf{Model}& \textbf{BLEU}& \textbf{CIDEr}& \textbf{ROUGE}& \textbf{SPICE}& \textbf{METEOR} \\ \midrule
            \textbf{Baseline 1}    &0.1942	&0.1342	&0.2527	&0.1028 &0.0929\\
            \textbf{Baseline 2}    &0.2122	&0.1635	&0.2748	&0.1654	&0.1394\\
            \textbf{Baseline 3}    &0.2093	&\textbf{0.1835}	&0.2377	&0.1555	&0.1407\\
            \textbf{Baseline 4}    &0.1953	&0.1798	&0.2471	&0.1478	&0.1407\\ \midrule[.01pt]
            \textbf{Proposed}      &\textbf{0.3020}	& 0.1817 &\textbf{0.3378}	&\textbf{0.1554}	&\textbf{0.1412}\\ 
            \bottomrule
        \end{tabular}
}}
\end{table}

%\noindent{\color{blue} Qualitative Results} qualitative evaluation
\subsubsection{Rank-based Evaluation}\label{sec:res_Qual_a}
The generated feedbacks are evaluated using MRR and Recall@k. As observed in Table~\ref{tab:qualres}, $76.58$\% feedbacks are relevant to one of the top $10$ comments and the MRR of $0.3789$ denotes that the generated feedbacks are contextually similar to one of the top $3$ comments. 

\begin{table}[!h] 
\centering
{%\fontsize{7}{10}\selectfont
    \caption{Rank-based Evaluation. `MRR' and `R@k' denote `Mean Reciprocal Rank' and `Recall@k' where k $\in$ \{1,3,5,10\}.}
    \label{tab:qualres}
    \resizebox{.42\textwidth}{!}{
        \begin{tabular}{@{}l|ccccc@{}}
            \toprule
            \textbf{Model}& \textbf{MRR}& \textbf{R@1}& \textbf{R@3}& \textbf{R@5}& \textbf{R@10} \\ \midrule
            \textbf{Baseline 1}    &0.3435		&17.30		&39.67		&60.67		&67.75 \\
            \textbf{Baseline 2}    &0.3305		&17.69		&36.99		&\textbf{61.47}  	&74.29 \\
            \textbf{Baseline 3}    &0.3214		&16.08		&37.53		&59.32  	&69.29 \\
            \textbf{Baseline 4}    &0.3182	 	&16.98	 	&37.26		&56.11 		&71.29 \\ \midrule[.01pt]
            \textbf{Proposed}      &\textbf{0.3789}	 	&\textbf{18.76}	 	&\textbf{40.92}		&60.13		&\textbf{76.58} \\ \bottomrule
        \end{tabular}
}}  
\end{table}

The variations in sentiment classification accuracy and MRR varied differently for different models. For example, baseline $4$ has lower MRR but high sentiment classification accuracy, whereas it is reverse for baseline $3$. The proposed model provides the right trade-off with high values for both.

\subsubsection{Sentiment-Control}
Table \ref{tab:control} reports the Control Accuracies, which represent the difference in accuracies between controlled and uncontrolled feedbacks. In controlled settings, the desired sentiment for the feedback to portray ($0$ for negative and $1$ for positive) is one of the input parameters. In uncontrolled settings, the parameter is not used and the control layer is disabled. %In contrast, controlled settings involve passing the ground-truth comment's sentiment as the parameter. 

\begin{table}[!h] 
\centering
\caption{Synthesized feedbacks' sentiment analysis. Here, `USentiAcc' and `CSentiAcc' denote the sentiment classification accuracies for uncontrolled and controlled feedbacks.}
\label{tab:control}
\resizebox{.42\textwidth}{!}{
\begin{tabular}{c|ccc}\toprule
 \textbf{Model}      &\textbf{USentiAcc} &\textbf{CSentiAcc} &\textbf{Control Acc}\\\midrule
 \textbf{Baseline 1}    & 52.34  &63.10  & 10.76\\
 \textbf{Baseline 2}    & 54.72  &67.06  & 12.34\\
 \textbf{Baseline 3}    & 48.25  &57.32  &  9.07\\
 \textbf{Baseline 4}    & 52.48  &71.57  & \textbf{19.09}\\ \midrule[.01pt]
 \textbf{Proposed}      &\textbf{58.41}	&\textbf{77.23} &18.82\\ 
 \bottomrule
\end{tabular}
} 
\end{table}	

The sentiment class of the feedback is determined using FLAIR \cite{akbik2019flair}, SentimentR \cite{sentimentr}, DistilBERT \cite{sanh2019distilbert}, and RoBERTa \cite{liu2019roberta}, as detailed in Section \ref{sec:dataset}. The model achieves a sentiment classification accuracy of $77.23\%$ and a control accuracy of $18.82\%$. To calculate the sentiment accuracy, one negative and one positive feedback are generated by passing the parameters $0$ for negative and $1$ for positive. The sentiment of the feedback is then calculated and compared to the ground truth sentiment labels. 

%\noindent\textbf{Sample Results}:
%Fig. \ref{fig:sample_res} shows sample results with uncontrolled, positively controlled and negatively controlled synthesized feedbacks.  

\subsubsection{Compute Time Requirement}
The total amount of compute time required for one epoch of model training on V$100$ GPU is as follows. Baseline $1$ required $168$ minutes, Baseline $2$ was the most time-consuming at $233$ minutes, Baseline $3$ needed $144$ minutes, and Baseline $4$ was the fastest at $103$ minutes. The Proposed System required $112$ minutes to complete one epoch.

\subsubsection{Interpretability and Visualization}\label{sec:inter}
\begin{figure*}
  \centering
  \includegraphics[width=.9\textwidth]{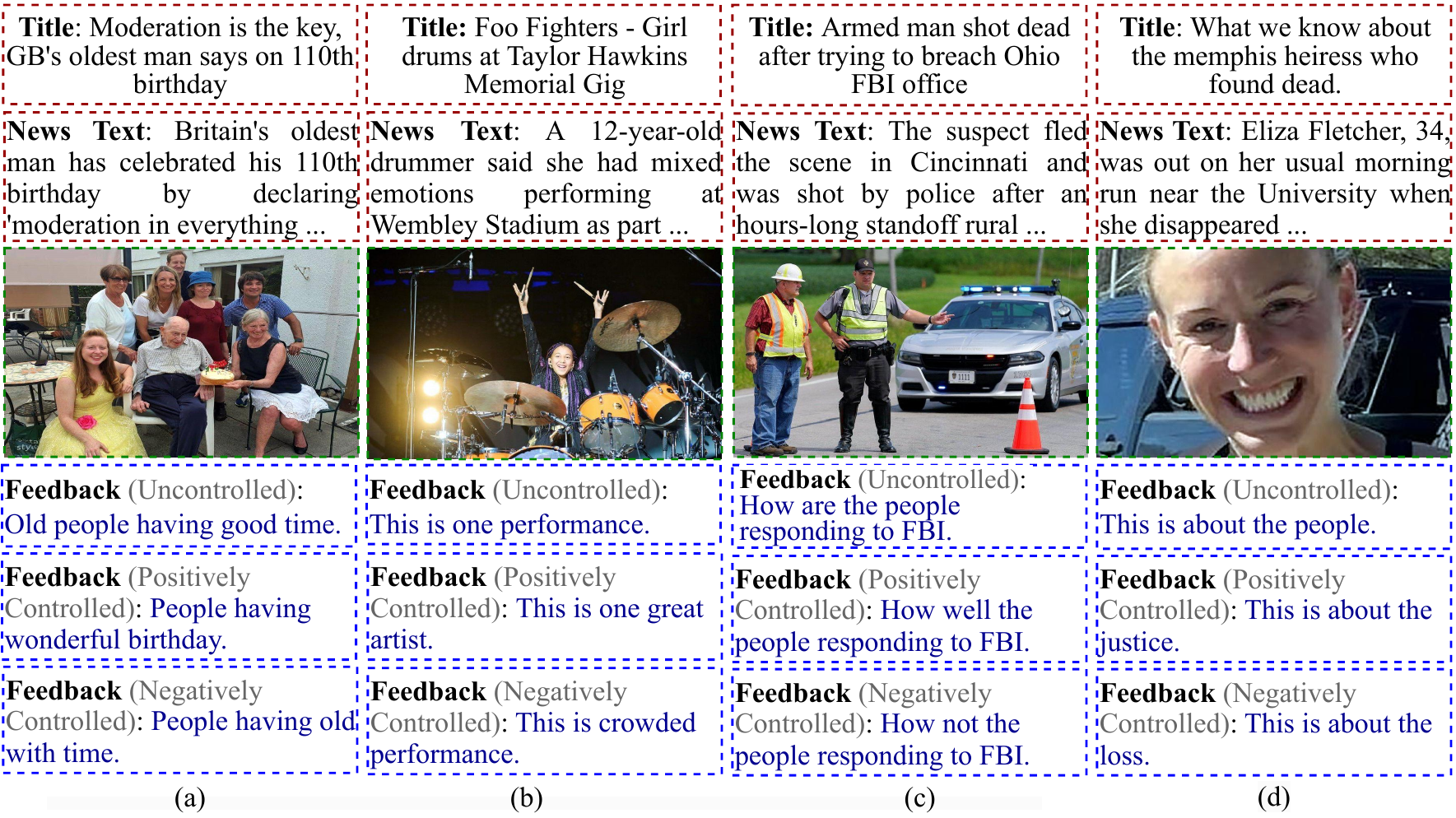}\vspace{-.1in}
  \caption{Sample feedbacks generated by the proposed system using input text and images (one out of multiple images shown) with sentiment-control. Supplementary material's Fig.~S1 depicts feature heatmaps, salient words and color denotations.} 
  \label{fig:sample_res} %
\end{figure*}

The model-level and case-level interpretability analyses have been incorporated. Model-level interpretability is achieved by introducing perturbations to the feedback synthesis model via the control layer. The impact of these perturbations on the output feedback is detailed in Table \ref{tab:control} in terms of sentiment classification accuracies for uncontrolled and controlled (perturbed) scenarios. Fig. \ref{fig:sample_res} shows sample results with uncontrolled and controlled feedbacks while Supplementary Sec.~S1 details pixel-level image attributions and word-level textual feature attributions.

\subsubsection{Human Evaluation} %\label{sec:human_eval}
The sentiments of the generated feedbacks were evaluated by $50$ evaluators ($25$ males, $25$ females; average age $30 \pm 2.73$ years). They assessed controlled and uncontrolled feedbacks for valence and relevance. A total of $50$ randomly picked samples were evaluated, and averages of evaluators’ scores were reported (Table \ref{tab:human_eval}). On average, $72.68$\% and $78.14$\% reported that positively and negatively controlled feedbacks are more positive and more negative, respectively, than uncontrolled feedbacks. Higher relevance scores for controlled feedbacks ($F_{PosCtrl}$, $F_{NegCtrl}$) than uncontrolled ($F_{UnCtrl}$) confirm the control layer’s influence on desired sentiment alignment. %Additional experimental results on interpretability analysis and ablation studies are included in Appendices \ref{sec:inter} and \ref{sec:ablation}, respectively.

\begin{table}[]
\centering	
\caption{Human evaluation of generated feedbacks where $F_{UnCtrl}$, $F_{PosCtrl}$ and $F_{NegCtrl}$ show uncontrolled, positively and negatively controlled feedbacks. $Rel_{img}$, $Rel_{text}$, $Rel_{Comment}$ and $Rel_{F_{UnCtrl}}$ are `relevant with' input images, text, comments and uncontrolled feedback, respectively.}
\label{tab:human_eval}
\resizebox{.5\textwidth}{!}
{%
    \begin{tabular}{c|cccc}\toprule
    \textbf{} & $\mathbf{Rel_{img}}$ & $\mathbf{Rel_{text}}$ & $\mathbf{Comment}$ & $\mathbf{Rel_{F_{UnCtrl}}}$\\ \midrule
    $\mathbf{Comment}$     &  70.85\% &   72.93\% &   100.00\% &  78.27\% \\
    $\mathbf{F_{UnCtrl}}$  &  67.27\% &   69.58\% &    78.27\% & 100.00\% \\
    $\mathbf{F_{PosCtrl}}$ &  69.47\% &   71.07\% &    79.93\% &  81.96\% \\
    $\mathbf{F_{NegCtrl}}$ &  71.23\% &   72.13\% &    80.17\% &  83.24\% \\ \bottomrule
    \end{tabular}
}\vspace{-.1in} 
\end{table}

\subsubsection{Ablation Studies}\label{sec:ablation}
Following ablation studies evaluate various parameters' impact on the proposed system's performance. 

\paragraph{Effect of Number of Control Layers and Value of Control-Parameter} The control layer is used after the decoder and before text generation to apply `control' on text generation. It is crucial to decide (i) the number of control layers and (ii) the value of the control parameter. We experimented with $1$, $2$, $3$, and $4$ control layers. The best performance was with $1$ layer, decreasing slightly with $2$, further with $3$, and significantly with $4$. For the control parameter, we tried $5$\%, $10$\%, $15$\%, and $20$\%. Increasing the value yields stronger control but trains fewer neurons on the full data, degrading quality. As shown in Table \ref{tab:ablation}, $10$\% gives the best trade-off. Hence, we use $1$ control layer with a control value of $10$\% in the final implementation. 

\paragraph{Effect of Beam-size} The beam size is a search parameter referring to the number of options the model keeps at each prediction step, controlling the breadth of the search for the best output. It keeps only the top $k$ predictions, where $k$ is the beam size. A larger beam size allows exploration of more possibilities, improving output quality, but increasing computation and causing repetitive text generation. We experimented with beam sizes $2$, $5$, $10$, $15$, and $20$; the corresponding sentiment classification accuracies and MRR values are shown in Table \ref{tab:ablation}. Beam size $5$ provided the best performance–complexity balance and is used in the final implementation.

\begin{table*}
\centering
{
    \caption{Ablation studies on control parameter, x and beam-size. The entries show sentiment classification accuracy/MRR.}
    \label{tab:ablation}
    \resizebox{.9\textwidth}{!}{
    \begin{tabular}{c|ccccc}
        \toprule
        \textbf{x / Beam-size}& \textbf{2}& \textbf{5}& \textbf{10}& \textbf{15}& \textbf{20} \\ \midrule
        \textbf{5}     &66.92\% / 0.3505 &76.89\% / 0.3605 &52.83\% / 0.3641 &54.51\% / 0.3214 &50.40\% / 0.3491 \\
        \textbf{10}    &71.42\% / 0.3483 &\textbf{77.23\% / 0.3789} &64.81\% / 0.3390 &50.48\% / 0.3393 &47.36\% / 0.3503 \\
        \textbf{15}    &52.82\% / 0.3429 &69.71\% / 0.3548 &54.99\% / 0.3312 &60.40\% / 0.3523 &45.24\% / 0.3449 \\
        \textbf{20}    &64.57\% / 0.3354 &75.12\% / 0.3389 &57.76\% / 0.3355 &49.63\% / 0.3409 &41.46\% / 0.3295 \\ \bottomrule
    \end{tabular}
}}
\end{table*}

\paragraph{Effect of Division Factor for KAAP technique} The suitable values of the division factors $k_{img}$ and $k_{txt}$ used in Section \ref{sec:interpretability} have been decided experimentally using the dice coefficient \cite{deng2018learning}. It measures two data samples' similarity; the value of $1$ denotes that the two compared data samples are completely similar, whereas a value of $0$ denotes their complete dis-similarity. For each modality, we computed the KAAP values at $k \in \{2, 3 \dotsc, 30\}$ and analyzed the dice coefficients for two adjacent $k$ values. For image \& text, the dice coefficient values converge to $1$ at $k$ of $5$ and $20$ respectively. Hence, the same have been used by the proposed system. 

%==================================
%-*-*-*-*-*- Conclusion -*-*-*-*-*-
%==================================
\section{Discussion and Conclusion}\label{sec:conclusion}
The proposed work's contributions, outcomes, limitations, and future directions for controllable multimodal feedback synthesis are summarized as follows.\vspace{.03in}

\noindent \textit{Advancements from Uncontrolled to Controllable Feedback Synthesis}: Building on our prior work on uncontrolled feedback synthesis \cite{kumar2023affective} utilizing basic data from Twitter without the comments metadata, we add sentiment control and interpretability. Using Facebook's relevance criteria and richer comment metadata, CMFeed improves ranking (MRR from $0.3042$ to $0.3789$ under comparable settings) and, trained on human comments plus multimodal inputs, generates sentiment-aligned feedback (Tables \ref{tab:quantre}, \ref{tab:qualres}). A control layer enables sentiment switching via targeted neuron activation.\vspace{.05in}

\noindent \textit{Ethical Considerations, Data Privacy, and Potential Misuse}:
All data collection follows Facebook’s policies and established ethical guidelines \cite{eth,eth2}. The process excludes personal identifiers and respects platform terms. The integrated interpretability module supports transparency and can help detect misuse, which reduces risks such as deceptive content and promotes responsible deployment.\vspace{.05in}

\noindent \textit{Generalizability and Diversity}:
Facebook’s user base of $3.07$ billion provides broad coverage and demographic diversity. The four selected news sources (BBC News, Sky News, Fox News, and NY Daily News) span different regions and political orientations. This choice reduces bias and enriches the range of perspectives represented in the dataset.\vspace{.05in}

\noindent \textit{Dataset Selection and Integration}:
We reviewed a Reddit dataset \cite{reddit_data} but did not include it because of unreliable sourcing, incomplete metadata, and weak ethical safeguards. We also chose not to merge Reddit or Twitter with Facebook to preserve consistent metadata, clear usage policies, and privacy controls.\vspace{.05in}

\noindent \textit{Limitations}:
The dataset may still contain platform-specific demographic biases. Training and deployment require significant computational resources. Adding audio or physiological modalities introduces additional privacy and ethics considerations. There is a risk of misuse for sentiment manipulation, which highlights the need for strict oversight and safeguards.\vspace{.05in}

\noindent \textit{Future Plans}:
We plan to incorporate discrete emotion classes to increase emotional granularity. We will explore additional modalities, such as audio and physiological signals, to enrich inputs. We also plan to control attributes such as emotion intensity and duration. Finally, we will conduct evaluations in education, healthcare, marketing, and customer service to assess effectiveness and ensure ethical compliance. 

% %========================================
% %-*-*-*-*-*- Acknowledgements -*-*-*-*-*-
% %========================================
\section*{Acknowledgements}
This work was supported by the University of Oulu and the Research Council of Finland Profi 5 HiDyn fund (grant 24630111132). The authors acknowledge the CSC-IT Center for Science, Finland, for providing computational resources.

% =*=*=*=*=*=*=*=*=*=*=*=*=*=*=*=*=*=*=*=*=*=*=*=*=*=*=*=*=*=*
\ifCLASSOPTIONcaptionsoff
  \newpage
\fi

% =*=*=*=*=*=*=*=*=*=*=*=*=*=*=*=*=*=*=*=*=*=*=*=*=*=*=*=*=*=*
% Ref section
\bibliographystyle{IEEEtran} 
\bibliography{ref}

\clearpage
\beginsupplement

\begin{strip}
\centering
\section*{\Large{Supplementary Material}\\~\\
Synthesizing Sentiment-Controlled Feedback For Multimodal Text and Image Data}
\end{strip}
\vspace{-1.25\baselineskip} % (optional) tighten gap before S1

%%%%%%%%%%%%%%%%%%%%%%%%%%%%%%%%%%%%%%%%%%%%%%%%%%%%%%%%%%%%%
\section{Detailed Interpretability Results}\label{suppl:inter}
Fig.~\ref{fig:sample_res_detailed} presents sample results illustrating model focus during generation for uncontrolled and controlled feedbacks. In image plots, red marks higher pixel contributions and blue marks lower contributions; in text plots, yellow marks higher word importance and blue marks lower importance. These visualizations reflect the KAAP attribution scheme: inputs are partitioned into \(k\) segments (image patches / word groups), SHAP-style contributions are estimated per segment, and then averaged to yield stable attributions. Thus, red/yellow regions correspond to segments with higher KAAP values. Comparing uncontrolled and controlled views isolates the effect of the control layer: when the sentiment mask switches, salient regions reweight while the input remains fixed. 

In Fig.~\ref{fig:sample_res1}, positive sentiment aligns with smiling faces and a family setting, while negative sentiment aligns with signs of ageing, especially in the older face. The left girl’s mixed expression (smile with discomfort) draws attention in both cases. The middle girl’s predominantly smiling face is highlighted in red for positive and in blue for negative. Fig.~\ref{fig:sample_res2} shows dark regions contributing to negative sentiment, with faces linked to positive sentiment. Under negative control, attention concentrates on the crowd; under positive control, it shifts to individual people. 

In Fig.~\ref{fig:sample_res3}, positive sentiment downplays the gun and emphasizes the number plate, while the uncontrolled setting focuses primarily on text. In Fig.~\ref{fig:sample_res4}, facial features are red under positive control and blue under negative control for the first image; in the second image, light and text associate with positive sentiment, whereas darkness associates with negative sentiment.

% ---- Single figure, split over two pages ----
\begin{figure*}[t]
    \centering
    % Page 1: (a) and (b)
    \subfloat[Sample Result 1\label{fig:sample_res1}]{
        \includegraphics[width=1\textwidth]{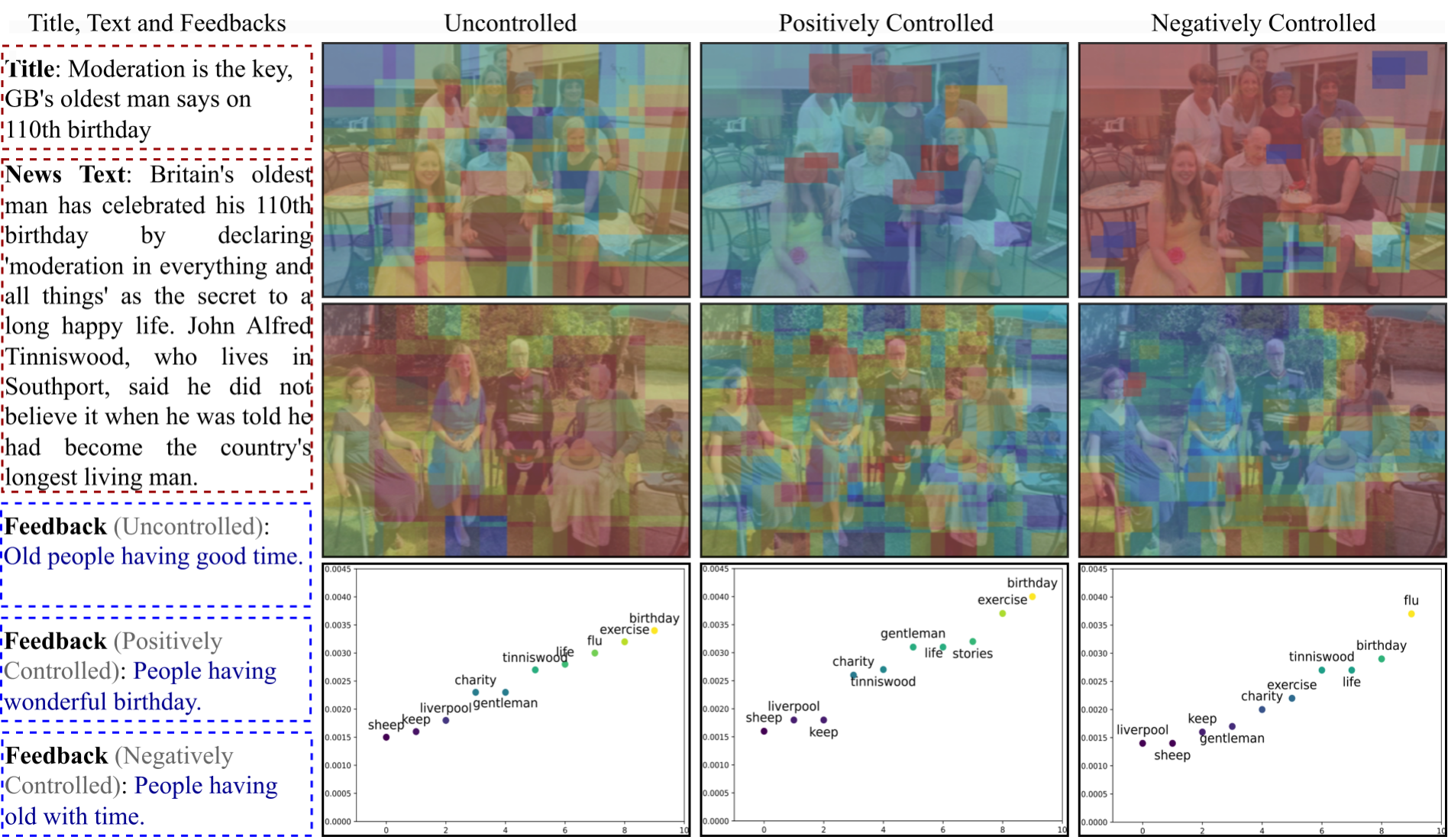}
    } 

    \subfloat[Sample Result 2\label{fig:sample_res2}]{
        \includegraphics[width=1\textwidth]{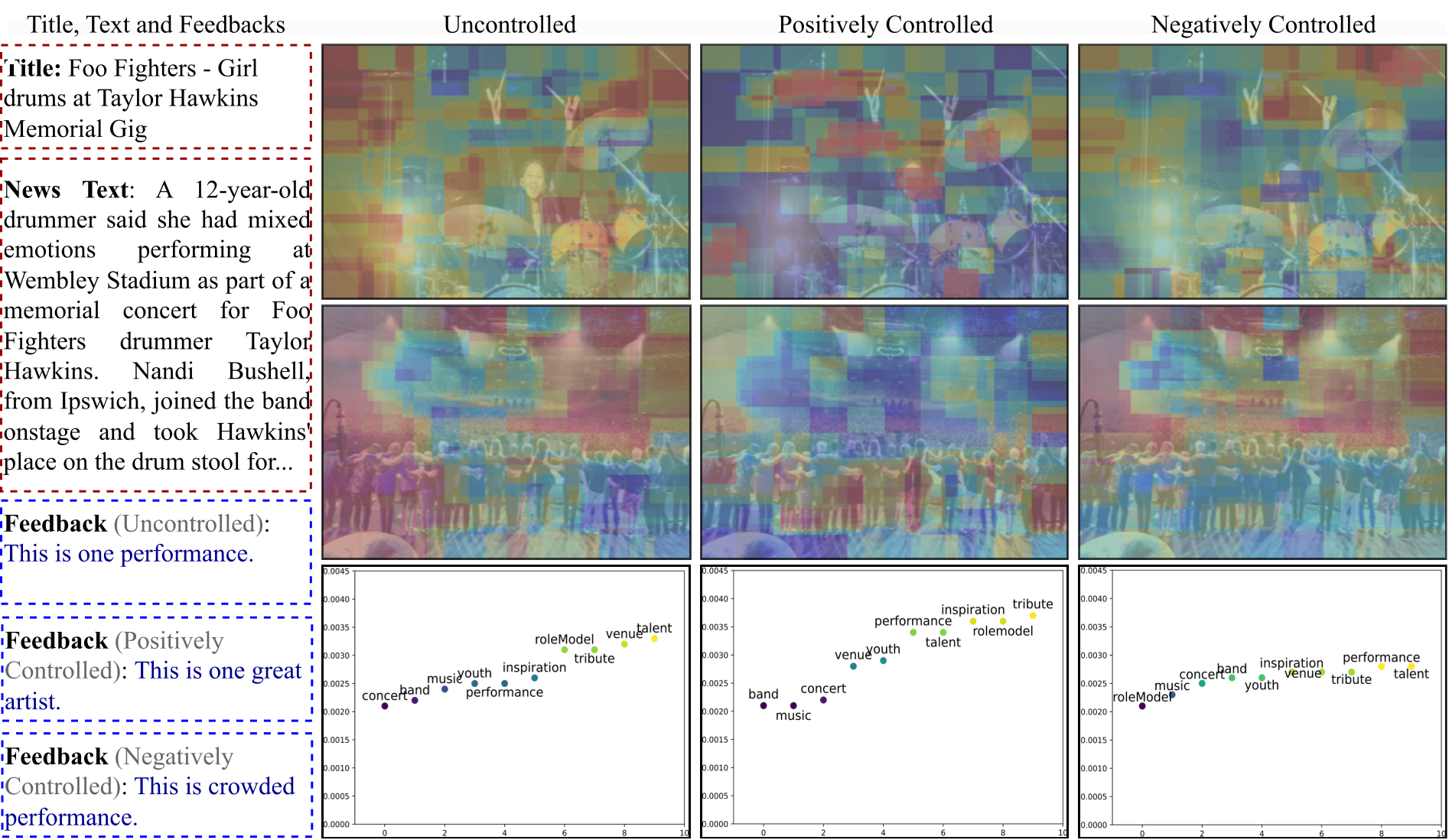}
    }\vspace{-.05in}
    \captionsetup{labelformat=empty}
    \caption{}\vspace{-2.0\baselineskip} 
    \captionsetup{labelformat=default}
\end{figure*}

\begin{figure*}[t]\ContinuedFloat
    \centering
    % Page 2: (c) and (d)
    \subfloat[Sample Result 3\label{fig:sample_res3}]{
        \includegraphics[width=1\textwidth]{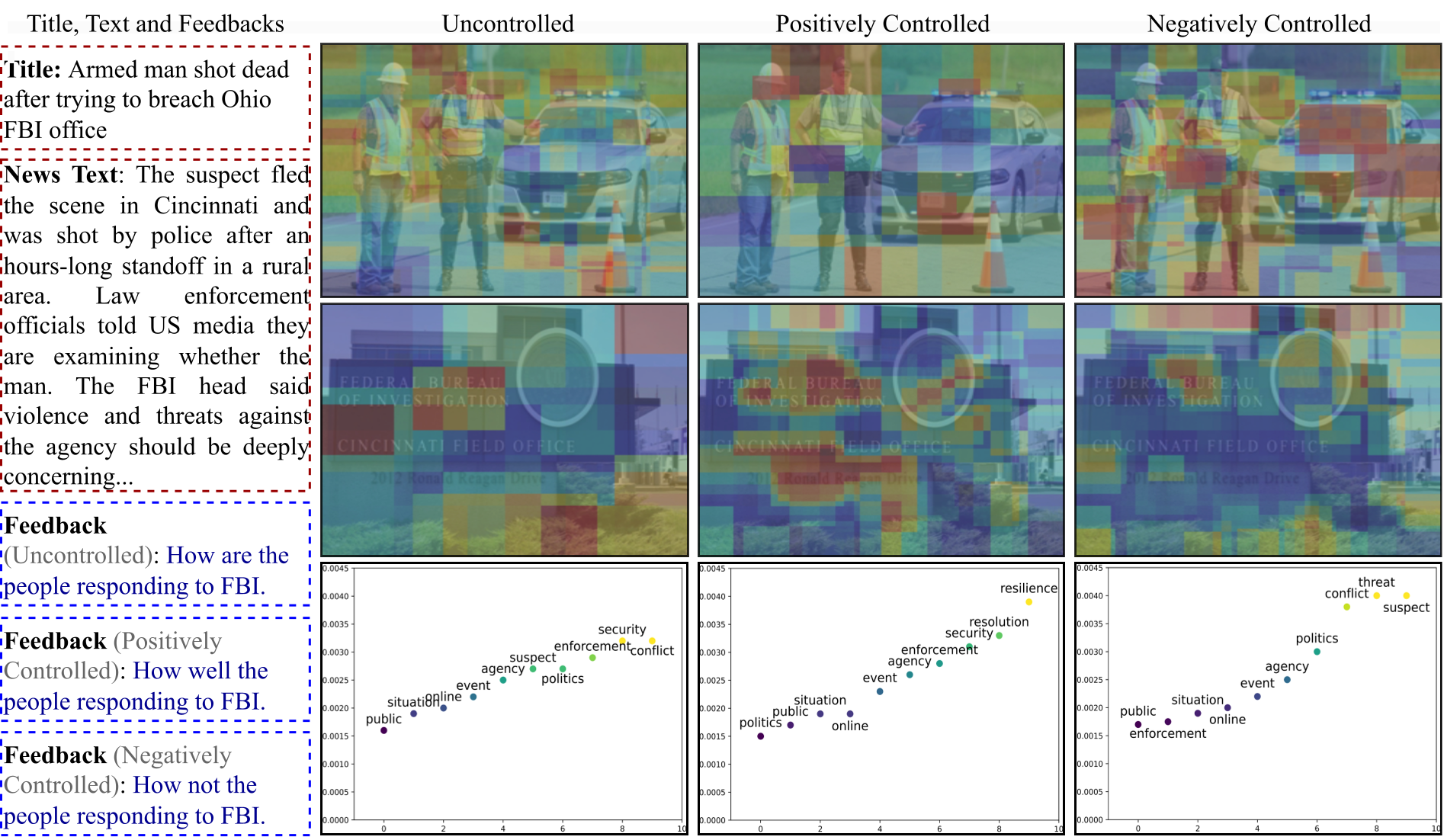}
    } 
    
    \subfloat[Sample Result 4\label{fig:sample_res4}]{
        \includegraphics[width=1\textwidth]{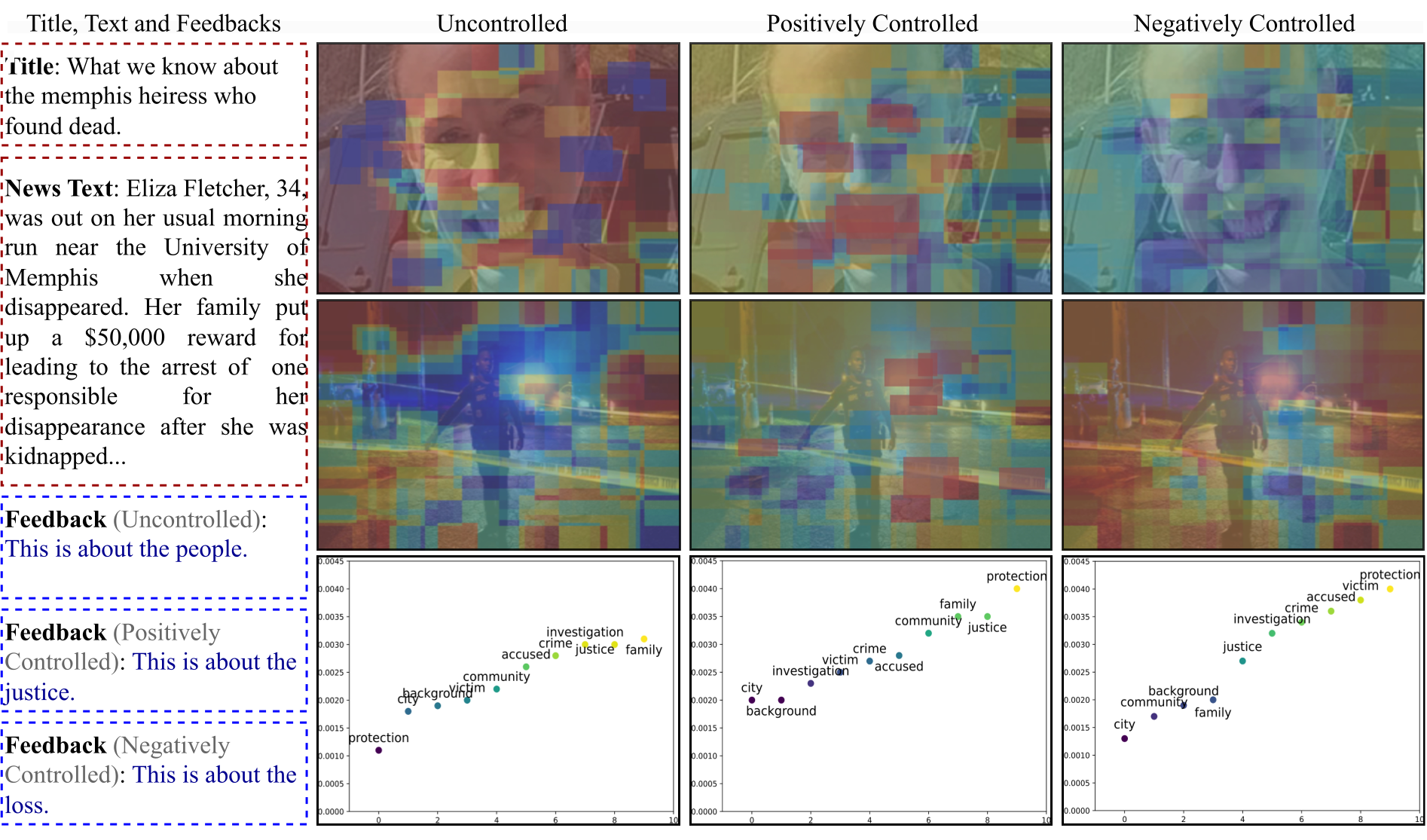}
    }\vspace{-.05in}
    \caption{Sample results along with interpretability plots. They depict the feedback generated by the proposed system using the news headline, text, and images (two out of multiple images shown) under the given sentiment-controllability constraint.}
    \label{fig:sample_res_detailed}
\end{figure*}

\section{Dataset: More Details} \label{suppl:dataset}
\subsection{Dataset \& Code Access, Documentation and Metadata} \label{data_details}
The CMFeed dataset is available for access at \href{https://zenodo.org/records/11409612}{zenodo.org/records/11409612} along with detailed documentation and metadata description. Its DOI is \textcolor{blue}{10.5281/zenodo.11409611}. The CMFeed dataset uses common formats: CSV for text and JPG for images. A detailed explanation of how to read and use the dataset is provided in this Zenodo repository. \newpage

The code for the controllable feedback synthesis models (for the baselines and proposed system) can be found at \href{https://github.com/MIntelligence-Group/CMFeed/}{github.com/MIntelligence-Group/CMFeed/} and used to reproduce the results. The steps for crawling data from Facebook posts are also described in this GitHub repository. \vspace{.12in}

\subsection{Licensing, Hosting, and Maintenance Plan} \label{license}
The CMFeed dataset has been hosted at the Zenodo repository, which is maintained by CERN's Data Centres. It ensures the long-term access and availability of the dataset. Under the `Creative Commons Attribution 4.0 International' license, the CMFeed dataset is released for academic research only and is free to researchers from educational or research institutes for non-commercial purposes.\vspace{.12in}

\subsection{Collection Process and Compliance with Ethical Standards}
Data for our study was systematically gathered using Facebook from publicly available posts on the pages of leading news channels such as Sky News, NYDaily, FoxNews, and BBC News. The data extraction process adhered to Facebook's terms and conditions outlined at \href{https://developers.facebook.com/terms/}{developers.facebook.com/terms/}, ensuring compliance with all applicable legal and ethical standards. Our data collection strategy supports responsible research, ensuring personal privacy and data security. Details of the data collection flow are provided in the aforementioned GitHub repository.\vspace{.12in}

\subsection{Intended Usage, Data Handling and Privacy}\label{privacypres}
The use of such data is intended solely for academic and research purposes, aiming to enhance the capabilities of feedback synthesis and sentiment analysis systems in a controlled environment.

The public nature of the collected data eliminates typical privacy concerns associated with personal user data. Throughout the data collection and handling stages, we rigorously filtered out any sensitive information, focusing on content that is inherently non-personal and intended for broad dissemination.

\end{document}